\documentstyle[aps,amssymb,epsfig,preprint]{revtex}
\newcommand{\R}{{\sf R\hspace*{-0.9ex}\rule{0.15ex}{1.5ex}\hspace*{0.9ex}}}

\begin{document} 
\title{A dissipative algorithm for wave-like equations in the
characteristic formulation}
\author{Luis Lehner}
\address{University of Pittsburgh, Pittsburgh, Pennsylvania 15260 \\
         e-mail: luisl@raven.phyast.pitt.edu}
\maketitle
\vspace{10cm}



\newpage

\noindent Proposed running head: Dissipative algorithm for wave-like equations. \\

\noindent {\bf Name:} Luis Lehner

\noindent {\bf Address:}\parbox[t]{8cm}{Center for Relativity, \\ The University of Texas at Austin,

Austin, TX 78712-1081,

USA}
			
\vskip2truemm
\vskip2truemm

\noindent {\bf Tel.:} 1-(512) 471-5426  

\noindent {\bf E-mail Address:}  luisl@feynman.ph.utexas.edu

\noindent {\bf Fax:} 1-(512) 471-0890

\newpage

\begin{abstract}
We present a dissipative algorithm for solving nonlinear wave-like
equations when the initial data is specified on characteristic
surfaces. The dissipative properties built in this algorithm make it
particularly useful when studying the highly nonlinear regime where
previous methods have failed to give a stable evolution in three dimensions.
The algorithm presented in this work is directly applicable to hyperbolic
systems proper of Electromagnetism, Yang-Mills and General Relativity theories.
We carry out an analysis of the stability of the algorithm and test its properties
with linear waves propagating on a Minkowski background and the scattering
off a Scwharszchild black hole in General Relativity.
\end{abstract}

\section{Introduction}
When modeling nonlinear problems, dissipative algorithms have provided
researchers with an extremely valuable tool since usually most non dissipative
schemes fail to produce a stable evolution. More precisely, when using
neutrally stable algorithms, instabilities often arise which spoil the
evolution. The addition of artificial
dissipation removes these instabilities by ``damping'' the
growing modes of the solution, in a somewhat controlled way. Therefore,
its inclusion in a discretization scheme provides a practical and
economic way of achieving longer evolutions.

The most widely used algorithms with this property are the family of
Lax schemes~\cite{thomas}; whereby adding to the equation $u_{,t}=- a
u_{,x}$ a term proportional to $u_{,xx}$ one  obtains a stable
discretization of the system that would otherwise be unstable.
However, one might correctly ask whether this is not tantamount to
solving a completely different problem. The beauty of these methods is
that the proportionality factor depends on the discretization size, and
in the continuous limit the approximation to modified PDE results in a
consistent approximation to the original one.

Although there is much experience with these kind of schemes, most of the
standard dissipative algorithms have been tailored for Cauchy initial
value problems, where initial data is provided at one instant of time
and evolved to the next instant by means of the
evolution equation.
However, in radiative problems, it is sometimes more convenient to
choose a sequence of hypersurfaces adapted to the propagation of the
waves, and therefore one adopts a foliation adapted to the
characteristics of the PDE under study.

In the present work, we present a new algorithm adequate for hyperbolic
systems. The underlying strategy of the proposed algorithm is quite
different from the conventional Cauchy-type methods. Rather, it is inspired by
analytic concepts developed in the 1960s~\cite{bondi,sachs,penrose} for
studies of gravitational radiation in General Relativity and in their
subsequent numerical integrations\footnote{For a
detailed account of the developments in the characteristic formulation
see~\cite{jeffnew}.}.
The main features of this approach are the use of characteristic
surfaces (for the foliation of the spacetime) and compactification
methods (that enable the inclusion of infinity in the numerical grid)
to describe radiation properties.
Although evolution algorithms (for systems possessing some kind of
symmetry) developed within this approach proved to be successful in the
linear and mildly nonlinear regime~\cite{papa,cce,dinv}, they produce
unstable evolutions when applied to the general case; which shows the need
for devising algorithms that could be applied in this situation.  In the
present work we present a new algorithm having dissipative properties,
making it a valuable tool to study systems where strong fields might
be present.

In section $2$ the details of the algorithm for the wave equation are
presented and its stability properties discussed. Section 3 is
devoted to treat a model problem which shows how the dissipative
algorithm might be a useful tool for numerical investigations in General
Relativity. Finally, in section $4$ we describe two particular
applications of this algorithm in the numerical implementation of
General Relativity.

\section{The algorithm}
Waves of amplitude $g$ traveling in one spatial direction with unit speed
obey the familiar equation
\begin{equation}
g_{,tt}-g_{,xx} = 0 ,
\label{weqcar}
\end{equation}  
Which can be solved in the region $R=\{ (t,x) \, / \, t \ge t_0,  x \in \R \} $,
assuming $g(t=t_0,x)$ and $g_{,t}(t=t_0,x)$ are given. If, instead, one
is interested in the solving the problem restricted to the region $x
\in [a, \infty)$, boundary data must also be provided corresponding to
$g(t,x=a)$. The analysis of this problem can be described in a simple
way in terms of the characteristics of this equation, which are given
by $(x-x_o)=\pm t$ through each spatial point $x_o$.

In particular, when solving Eq. $(\ref{weqcar})$ in the region $\cal
R_C$. The domain of dependence ${\cal D_C}$ of a point $(t_1,x_1)$
is given by ${\cal D_C} = \cal S_C \bigcap \cal R_C$, with $\cal S_C$
naturally defined by the characteristics passing through $(t_1,x_1)$ as
\begin{equation}
{\cal S_C} = \{(t,x) \mbox{ such that } t \le t_1 \mbox{ and }
 (t-t_1)^2 - (x-x_1)^2 \ge 0 \} \, ;
\end{equation}
and $\cal R_C$ is the region to the future of
\begin{itemize}
\item{the line $t=t_0$}, \\  \mbox{                     } {\bf or}
\item{the region defined by  $[a,\infty)$ or $x \in [a, b]$  (where
$a \in \R$); in these cases, boundary conditions must be imposed at 
$x=a$ (and $x=b$ in the latter case).}
\end{itemize}

Suppose one introduces a coordinate system adapted to the characteristics
by, say $(u=t-x, r=x)$; then, Eq. (\ref{weqcar}) reduces to
\begin{equation}
2g_{,ur}-g_{,rr} = 0. 
\label{weqnul}
\end{equation}
Further, one can then choose to foliate the spacetime by a sequence of
characteristics obtained by holding the (retarded) time $u=const$. One can
then define a {\it characteristic initial value problem}, where Eq.
(\ref{weqnul}) is solved provided that $g(u=u_o,r)$ is given. (Note
that $g_{,u}(u=u_o,r)$ need not be provided as in the Cauchy initial
value problem).

It is straightforward to check that a solution of Eq. (\ref{weqnul}) is
expressible as $g(u,r) = F(u)+G(u+2r)$ (where $F$ and $G$ are smooth
functions).  Physically, $F(u)$ describes waves propagating in the $+
r$ direction (outgoing waves) and $G(u+2r)$ describes waves propagating
in the $- r$ direction (incoming waves). Then, if one imposes the
condition of pure outgoing waves, the solution must be of the form
$g=F(u)$; hence, along each characteristic the value of the function is
constant. Notably, boundary data at $r=0$ can be given arbitrarily
since purely outgoing waves at $u=u_0$ will not reach $r=0$. More generally,
boundary data consistent with the incoming waves must be prescribed at
$r=0$.

It is important to note the domain of dependence
for this problem. When solving Eq. (\ref{weqnul}) in the region
${\cal R}_c$, the domain of dependence $({\cal D}_c)$ of a point $(u_1,
r_1)$ is defined by ${\cal D}_c = {\cal S}_c \bigcap {\cal R}_c$
where
\begin{equation}
{{\cal S}_c} = \{(u,x) \mbox{ such that } u \le u_1
 \mbox{ and } (u-u_1)^2 + 2 (u-u_1) (r-r_1) \ge 0 \}.
\end{equation}
However, if the region ${\cal R}_c$ is chosen to be the
future of the line $u = u_0$, ${\cal D}_c$ extends indefinitely to the
past.
Therefore, the characteristic approach requires ${\cal R}_c$ to have a
boundary. Thus, one defines ${\cal R}_c$ as the region $( u \ge u_o, r
\in [a, \infty) )$ (with $a \ge 0$). Figure (\ref{fig:domdep})
illustrates the domains of dependence corresponding to each
formulation.

For hyperbolic systems with two or more spatial dimensions, the manner 
in which the characteristics determine the domain of dependence and lead
to evolution equations is qualitatively the same. Also, the use of
coordinates adapted to them provide a tidy way for studying the
system. For instance, in $3$ dimensions,
the wave equation is given by
\begin{equation}
\Psi_{,tt} - \Psi_{,xx} - \Psi_{,yy} - \Psi_{,zz} = 0 \, ,
\end{equation}
which, in term of spherical polar coordinates $(t, r ,\theta, \phi)$
has the form
\begin{equation}
r \Psi_{,tt} - (r \Psi)_{,rr} - L^2 \Psi /r = 0 \, ,
\label{wav3d}
\end{equation}
where $L^2$ denotes the angular momentum operator
\begin{equation}
L^2 \Psi =  \frac{(\sin(\theta) \Psi_{,\theta})_{,\theta}}{\sin(\theta)}
+ \frac{\Psi_{,\theta \theta}}{\sin^2(\theta)} \, .
\end{equation}
Introducing coordinates $(u=t-r, r, \theta, \phi)$, which defines a natural
inner boundary at $r=0$, Eq. (\ref{wav3d})
takes the form
\begin{equation}
2 (r \Psi)_{,ur} - (r \Psi)_{,rr} = \frac{L^2 \Psi}{r} \, .
\label{3d1d}
\end{equation}
Thus, by defining $g \equiv r \Psi$ and considering $L^2 \Psi/r$ as a source
term, Eq. (\ref{3d1d}) formally looks like the $1$ dimensional system. Therefore,
from now on we restrict our analysis to this latter case and extend our
results to the $3$ dimensional case in section $4$.

The formal integration of (\ref{weqnul}) proceeds by an integration in
the $r$ direction on each $u=const$ surface and then evolve to the next
level. This reformulates the integration in the
characteristic formulation as an ``evolution'' in the
radial direction and then another in the $u$ direction (as opposed to the
evolution of a ``whole'' instant of time to the next one typical of
the Cauchy evolution). Hence, standard dissipative schemes intended for
Cauchy-type evolutions (like the family of Lax algorithms) are not
directly applicable in the characteristic formulation of the PDE and the
addition of artificial viscosity to the system must be reformulated.

In the numerical implementation of Eq. (\ref{weqnul}) a useful
discretization was introduced in~\cite{isaac}. This scheme is basically
a second order approximation based on finite difference techniques.
Assuming the grid discretization is given by $u_n = n \Delta u$ and $r_i
= i \Delta r$, the derivatives may be discretized in the following way:
\begin{eqnarray}
{g_{ur}|}^{n+1/2}_{i-1/2} &=& 
\frac{g^{n+1}_i-g^{n+1}_{i-1}-g^n_i+g^n_{i-1}}{\Delta u \Delta r} \label{gur} \, , \\
{g_{rr}|}^{n+1/2}_{i-1/2} &=& \frac{g^{n+1}_i-2g^{n+1}_{i-1}+g^{n+1}_{i-2} +
		  g^{n}_{i+1}-2g^n_i+g^n_{i-1}}{\Delta r^2} \label{grr} \, .
\end{eqnarray}
The resulting scheme (which we will refer to as GIW) is a second order
in time, second order in space accurate algorithm. Notably, the Von Neuman
analysis shows that the GIW scheme has unitary amplification factor (i.e. a
neutrally stable algorithm) independent of the values of $\Delta u$ and
$\Delta r$. This would imply that the algorithm is unconditionally
stable which is at first sight puzzling.  This might be explained by
the implicit local structure of the algorithm (since it involves $3$ points
at the upper time level) and, as such, a local stability analysis need
not give a condition on the discretization size. Nevertheless, the
algorithm is globally explicit as the evolution proceeds by an outward
march from the origin. Hence, the algorithm does require the
enforcement of the CFL condition to ensure that the numerical and analytical
domains of dependence are consistent.

The CFL condition for the system can be easily obtained. The field at
grid point at $(u_1,r_1)$, depends critically on the field value at
$(u_1-\Delta u,r_1 + \Delta r)$ (since all the points where $0 \le r
\le r_1$ are trivially included in the discretization). The requirement for 
the numerical domain of dependence to include the analytical domain of
dependence is $ \Delta u^2 - 2 \Delta u \Delta r \le 0$; therefore, the CFL
condition will be satisfied if $\Delta u \le 2 \Delta r$.

The GIW algorithm has been employed successfully in the characteristic
formulation of General Relativity (G.R.) assuming either spherical
symmetry~\cite{isaac,excision}; axisymmetry~\cite{papa} or very small
departures from spherical symmetry~\cite{cce}. However, when
considering more general problems, as it is often the case with neutrally
stable schemes, round off error or parasitic modes are enough to cause
ripples in the solution which often lead to an unstable evolution. As
stated earlier, adding dissipation to the PDE constitutes a way to
alleviate this problem~\cite{thomas}. We now show that
 a rather simple modification of (\ref{weqnul}) can be used to
obtain a consistent discretization with dissipative properties.

We start by considering the following equation 
\begin{equation}
2 g_{,ur}- g_{,rr} + 4/3 \epsilon \frac{\Delta r^2}{2 \Delta u} g_{,rrr} = 0;
\label{neweqn}
\end{equation}
(the $4/3$ factor is included for convenience).
A straightforward discretization of (\ref{neweqn}) is obtained by the
described approximation for $g_{,ur}$ (\ref{gur}) and $g_{,rr}$
(\ref{grr}) and by approximating the third derivative at the point
$(n,i-1/2)$ as
\begin{equation}
{g_{,rrr}|}^n_{i-1/2}=\frac{1}{\Delta r^3}
(g^n_{i+1}-3g^n_{i}+3g^n_{i-1}-g^n_{i-2}) \, .
\end{equation}

In analogy to the Lax method, the inclusion of this extra term leads to
a consistent difference approximation of Eq. (\ref{weqnul}); that
is, the difference approximation converges formally to the differential
equation in the limit $(\Delta u, \Delta r) \rightarrow 0$. In fact, it is
straightforward to check that the resulting approximation is accurate of
order $\{ {\cal O}(\Delta r^2), {\cal O}(\epsilon \Delta t) \}$\footnote{Contrary
to the Lax-method which exhibits strict second order convergence in
space and time.}. An important
feature of the resulting algorithm (which we shall call DA) is its
dissipative features, which make it particularly useful. The stability
properties of this algorithm can be easily obtained by introducing Fourier
modes such that $g=e^{su} e^{ik j/\Delta r}$. After some algebra one
obtains
\begin{equation}
S \bigg( i + 2 \alpha \sin(k\Delta r/2) e^{-ik\Delta r/2} \bigg) = 
i \bigg( (1-\epsilon) + \frac{4}{3}\epsilon ( 4 \cos^2(k\Delta r/2) - 1)\bigg) - 2 \alpha \sin(k\Delta r/2) e^{-ik\Delta r/2} \, ,
\end{equation}
where $S \equiv e^{su}$ and $\alpha = \Delta u /(4 \Delta r)$.
Therefore, the equation governing the growth of the solution's modes is
\begin{equation}
|S|^2 = 1 + \frac{4 \epsilon \sin^2(k\Delta r/2)}{3 (1-
4\alpha (1-\alpha) \sin^2(k\Delta r/2) )} \bigg(-2 + \sin^2(k\Delta r/2)
(4 \alpha + 4/3 \epsilon ) \bigg)
\end{equation}
Now, since $4\alpha (1-\alpha) \sin^2(k\Delta r /2) < 1$ (for $\alpha <
1/2$) the scheme will be stable if, $0 \le \epsilon \le 3/2 (1 - 2 \alpha)$.
Moreover, $ |S| < 1$ as $k \rightarrow \pi/\Delta r$, indicating that
the high frequency modes are effectively ``damped'', while $|S| \rightarrow 1$ 
as $k \rightarrow 0$. 
 
The obtained discretization can also be thought of as an approximation
to the original equation (\ref{weqnul}) (i.e. without the addition of
the extra term) where the finite differencing of $g_{,ur}$
includes $4$ points on the {\it nth} level as
\begin{eqnarray}
{g_{ur}|}^{n+1/2}_{i-1/2} &=& \frac{{g_{,r}|}^{n+1}_{i-1/2} -
 {g_{,r}|}^{n}_{i-1/2}}{\Delta u} \nonumber \\
&=& \frac{g^{n+1}_i-g^{n+1}_{i-1}}{\Delta u \Delta r} - 
 \frac{(1 - \epsilon) (g^n_{i}-g^n_{i-1})  +
       \epsilon (g^n_{i+1}-g^n_{i-2})/3 }{\Delta u \Delta r}\, ,
\end{eqnarray}
 which can be regarded as a weighted
average of the derivatives at $(n,i-1/2)$ obtained from field
values at the points $\{ (n,i), (n,i-1) \} $ and $ \{ (n,i+1), (n,i-2) \}$. 
In the next section, we illustrate
how this algorithm produces a stable discretization when the original
strategy (corresponding to $\epsilon=0$) fails.

\section{Application in a ``toy problem''}
In this section we study the stability properties of an equation bearing
close resemblance to the nonlinear evolution equation encountered in the
characteristic formulation of General Relativity (which will presented in the section IV)
\begin{equation}
  2\, G_{,ur} - G_{,rr} = G \, G_{,u} \, G_{,r}.
  \label{eq:biggnon}
\end{equation}
In order to keep track of the nonlinearity of the equation, we
introduce the parameter $\lambda$ (with $\lambda \leq 1$),
such that $G \equiv \lambda g$; hence,
Eq.~(\ref{eq:biggnon}) becomes
\begin{equation}
  2\, g_{,ur} - g_{,rr} = \lambda^2  g \, g_{,u} \, g_{,r}.
  \label{eq:biggnongral}
\end{equation}

In particular, note that the principal part of Eq.
(\ref{eq:biggnongral}) corresponds to the wave equation. Also, it
reduces in the  the linear case ($\lambda^2=0$), to the wave equation.
Consequently; one might expect the GIW 
discretization to lead a stable scheme.

However this is not the case, as can be demonstrated by the
following analysis. First, in order to simplify the study of the
stability of this nonlinear problem, we {\it linearize the
PDE with respect to the previous time
step}\cite{thomas} to obtain a more manageable equation. In
this linearization, we approximate the values of $g$ and $g_{,r}$ with
respect to the {\it nth} level, but $g_{,u}$ is centered in between the
levels. The resulting finite difference approximation is
\begin{eqnarray}
 \lefteqn{ g_{i}^{n+1}-g_{i-1}^{n+1}-g_{i}^n+g_{i-1}^n } \nonumber \\
  & & \mbox{}+ { {\Delta u} \over {4 \Delta r}}
 \left( - g_{i}^{n+1} + 2\, g_{i-1}^{n+1} - g_{i-2}^{n+1}
        - g_{i+2}^n     + 2\, g_{i}^n - g_{i-1}^n \right) =  
 \nonumber \\
 & & \mbox{} \; \; \lambda^2 \; \frac{1}{8}(g_{i}^n + g_{i-1}^n) (g_{i}^n - g_{i-1}^n) 
    ( g_{i}^{n+1} + g_{i-1}^{n+1} - g_{i}^n - g_{i-1}^n)\, .
\end{eqnarray}
Finally, we introduce the Fourier modes $g=e^{s u}e^{ik j\Delta r}$ and
solve for $|S|^2$, obtaining
\begin{eqnarray}
|S|^2 = 1 
+ \frac{16 \alpha \lambda^2 \sin(K)^2 (1+\cos(K))}{D} \label{eq:oldsta}\, ,
\end{eqnarray}
where $\alpha = \Delta u/(4 \Delta r) $,  $K=k \Delta r$ \, and
\begin{eqnarray}
D \equiv & & 16 + \lambda^4 (1+\cos(K))^2 - 8 \lambda^2 \bigg( \alpha
\sin^2(K) + \cos(K) (1 +\cos(K)) \bigg)\nonumber \\ & & + 32 \alpha
(1-\alpha) (\cos(K)-1) \label{denom}\, .
\end{eqnarray}
It is not difficult to check that $D$ is a positive quantity for 
\begin{equation}
0 < \alpha \leq \frac{1}{2} - \frac{\lambda^2}{8} + 
\frac{\sqrt{48 + \lambda^4 - 40\lambda^2}}{8}
\end{equation}
(which will be the case if the CFL condition is satisfied).
Thus, the value of $|S|$ is always larger than $1$, indicating that this
discretization is {\it unconditionally unstable!} It is remarkable that
the simple addition of some nonlinear terms, even when they do not
change the equation's principal part, completely break an algorithm that
would otherwise be stable. 

We now modify the way $g_{,ur}$ is discretized by introducing dissipation as
dictated by the D.A. scheme, i.e.
\begin{equation}
g_{,ur} = \frac{1}{\Delta u \Delta r} 
 \bigg(g_{i}^{n}-g_{i-1}^{n}- ((1-\epsilon) g_{i}^n - g_{i-1}^n 
 + \epsilon(g_{i+1}^n - g_{i-2}^n)/3) \bigg)\, .
\end{equation}
With this simple modification, the value of $|S|^2$ now becomes
\begin{eqnarray}
|S|^2 = 1 + \frac{4 (\cos(K)-1) N}{D} ,
\end{eqnarray}
where
\begin{eqnarray}
N &=& - 4 \alpha \lambda^2 (1+\cos(K))^2 \nonumber \\
&& +  \epsilon  \bigg( 4 + \epsilon (\cos(K)-1)   
 +\lambda^2 \cos(K) (\cos(K) + 1) + 4 \alpha (\cos(K) -1) \bigg)  
 \label{eq:cond1} .
\end{eqnarray}
Since $D > 0$ and $\cos(K)-1 \le 0 $, the condition for stability is
that $N \ge 0 $. Given $\alpha$, this a condition on $\epsilon$, or
vice versa. For instance, if $\alpha=1/8$ and $\lambda = 2^{-1/2}$, the
discretization will be stable if $0.3 \le \epsilon \le 1$.  On the
other hand, if we choose $\lambda = 1$ and $\epsilon = 1/4$, $N \ge 0$
will be satisfied if $\alpha \leq \frac{1}{8}$. Figure~\ref{vareps}
shows the value of $|S|$ for different choices of $\epsilon$ for a
given $\lambda$; the effect of the added dissipation can be 
clearly seen.

\section{A practical application. The characteristic formulation of G.R.}
When solving Einstein equations, one can take advantage of the coordinate
invariance of the theory to simplify the modeling of a specific
problem. In particular, one is free to choose a foliation of the
spacetime better suited to the problem.

In the numerical implementation of G.R., the most common approach is to
choose a foliation by spacelike  hypersurfaces at constant times. In this
 approach, Einstein
equations form a second oder PDE system for the {\it intrinsic
geometry} of each surface and its embedding in the spacetime, the {\it
extrinsic curvature}.  Einstein equations are split into two distinct
set of equations. One set consists of {\it constraint} equations that
limit the possible configurations of the field variables on each
hypersurface. The second set constitutes the {\it evolution} equations
that determine the development of the field variables onto the next
hypersurface\footnote{For a complete description of this formalism
see~\cite{mtw}.}.

The main drawback of the numerical implementation of the Cauchy
formulation is the impossibility of having an infinite grid to
completely cover the spacelike hypersurfaces. Thus, in practice, one
chooses an exterior boundary in order to deal with a finite domain.
This introduces further problems since special conditions on the
boundary must be imposed in order to avoid reflections which spoil long
term evolutions. Although in the $1$ dimensional case there are sound
method to achieve this (e.g. the  Sommerfeld condition), in the
general case, any local boundary condition still introduces reflections
turning the task of obtaining long accurate evolutions into an almost
impossible one. A related problem arising from an outer boundary at a finite
distance is that the radiation can not be rigorously calculated.

When studying gravitational radiation a more natural choice adapted to the wave
propagation is to adopt a sequence of characteristic hypersurfaces to cover the
spacetime. This approach is known as the {\it characteristic formulation of G.R.},
pioneered by Bondi and Sachs~\cite{bondi,sachs}. The main ingredients of this
formulation are the foliation of the spacetime by a sequence of characteristic
hypersurfaces and the use of compactification techniques (which enable the inclusion of
infinity in a finite grid) to rigorously describe asymptotic properties of
radiation~\cite{penrose}. The equations naturally split into {\it hypersurface
equations} and {\it evolution equations}. We now outline the main aspects of the
numerical implementation of this formulation (based on~\cite{cce,hpgn}; where
a detailed description of the problem has been presented) and employ the constructed
algorithm to discretize the PDE equations governing the evolution of the fields.

A coordinate system is introduced by labeling the outgoing lightlike hypersurfaces with a
parameter $u$; ach null ray on a specific hypersurface labeled with $x^A$ $(A=2,3)$ and choose
 $r$
as a surface area coordinate (i.e. surfaces at $r=const$ have area $4 \pi r^2$). In the
resulting $x^a=(u,r,x^A)$ coordinates, the metric takes the Bondi-Sachs form~\cite{bondi,sachs}
\begin{eqnarray}
   ds^2 & = & -\left(e^{2\beta}V/r -r^2h_{AB}U^AU^B\right)du^2
        -2e^{2\beta}dudr -2r^2 h_{AB}U^Bdudx^A 
         +  r^2h_{AB}dx^Adx^B,    \label{eq:bmet}
\end{eqnarray}
where $h^{AB}h_{BC}=\delta^A_C$ and $det(h_{AB})=det(q_{AB})$, with
$q_{AB}$ a unit sphere metric. 

The metric components are re-expressed as
\begin{eqnarray}
h_{22}&=&\frac{4}{P^2}(\Re[J] + K),\;\nonumber \\
h_{23}&=&h_{32}=\frac{4}{P^2}\Im[J],\;\nonumber \\
h_{33}&=&\frac{4}{P^2}(K-\Re[J]), \nonumber \\
U^2&=&\frac{P}{2}\Re[U],\; \nonumber \\
U^3&=&\frac{P}{2}\Im[U];
\label{eq:huq}
\end{eqnarray}
where $P=\sec^2(\theta/2)$ in standard angular spherical coordinates
$(\theta, \phi)$. Here, the metric is expressed in terms of two
real ($\beta$ and $V$) and two complex ($U$ and $J$)
variables (where
$K=\sqrt{ 1+J \bar J}$).  The complex field $J$ measures the departure
of spherical symmetry of the surfaces given by $r=const$, and
$u=const$; $V$ represents the mass distribution of the system; $\beta$ measures
the expansion of the light rays and $U$ measures the shift in the angular
directions from one hypersurface to another (at constant $r$).

The hypersurface equations are expressed as:
\begin{eqnarray}
\beta_{,r} &=& F_{\beta}[J] \label{beta} \\
U_{,r}  &=& F_U[\beta,J] \label{u} \\
(r^2 Q)_{,r}  &=& F_Q[U,\beta,J] \label{q} \\
V_{,r}&=& F_V[Q,U,\beta,J], \label{w}
\end{eqnarray}
where $Q \equiv r^2 e^{-2\beta} (J \bar U_{,r} + K U_{,r}) $ which is introduced
to deal with a first order system of hypersurface equations. The functions
$F_{\beta}, F_U, F_Q$ and $F_V$ involve derivatives taken only on a particular hypersurface
${\cal N}$. Then, they can be easily integrated if $J$ is known on ${\cal N}$ (assuming
consistent boundary conditions are provided) in the following
way. The integration strategy proceeds by first obtaining $\beta$ from Eq.~(\ref{beta}), 
then $U$ from Eq.~(\ref{u}), followed by the calculation of $Q$ 
using Eq.~(\ref{q}) and
finally $V$ using Eq.~(\ref{w}). The evolution to the next hypersurface is prescribed
by a first order (in time) equation for $J$ that takes the form
\begin{eqnarray}
    2 \left(rJ\right)_{,ur}
    - \frac{V}{r}\left(rJ\right)_{,rr} &=& \, r \, J  \bigg ( \frac{J_{,u}}{K} 
    (\bar J_{,r} K - \bar J K_{,r} ) + c.c. \bigg) + F_J[V,Q,U,\beta,J]
    \label{eq:wev}
\end{eqnarray}
where $F_J$ involve derivatives on ${\cal N}$ only.

A code that implements Einstein equations was written using
(second order) finite difference approximations. Angular and radial
derivatives are approximated along the following lines~\cite{hpgn}, 
\begin{itemize}
\item{{\it Angular derivatives}. 
We follow the formalism given in~\cite{newt,nullinf}.
To expedite the numerical implementation of angular derivatives, instead
of working with the standard spherical angular coordinates $(\theta,
\phi)$, we work in stereographic coordinates:
\begin{equation}
 x^A=(q,p)=(\tan(\theta/2) \cos(\phi), \tan(\theta/2) \sin(\phi) )\, , \nonumber
\end{equation}
and angular derivatives are written in terms of the (complex
differential) eth operators $\eth$ and $\bar \eth$~\cite{goldberg,bms}; for instance,
\begin{equation}
\frac{\partial \beta}{\partial q} = \frac{ \eth \beta + \bar \eth \beta}{P} \, .
\end{equation}
This allow us to employ a set of numerical techniques introduced in~\cite{eth} which are
specially tailored to: ({\it i}) handle the numerical approximation of angular derivative
operators and ({\it ii}) deal with the fact that a single coordinate patch can not be used to
smoothly cover the sphere. }
\item{{\it Radial derivatives}. These are approximated via centered second order
differences along each null ray (i.e. holding $x^A=const$); for instance
\begin{equation}
\beta^n_i = \beta^n_{i-1} + \Delta r F_\beta|^n_{i-1/2} \, .
\end{equation}
}
\end{itemize}

The evolution equation deserves special consideration. Its discretization (in
between levels $n$ and $n+1$) is
obtained using dissipation in the following way: schematically, it
can be re-expressed as
\begin{equation}
  2\, g_{,ur} - (V/r) g_{,rr} = \frac{g}{r^2} \, ( \bar g_{,u} \, g_{,r} + c.c.) +
F_J[\beta, J , U, V]. \, ,
  \label{eq:evol}
\end{equation}
where $g \equiv rJ$. The function $F_J$ can be straightforwardly approximated
at each grid point $(n+1/2,i-1/2)$ to second order accuracy. Then, in
order to introduce dissipation in the algorithm, we proceed to consider
a modified version (along the lines described in section II)
\begin{equation}
  2\, g_{,ur} - (V/r) g_{,rr} = \frac{g}{r^2} \, ( \bar g_{,u} \, g_{,r} + c.c.) +
F_J(\beta, J , U, V) + \epsilon \frac{\Delta r^2}{\Delta u} g_{,rrr}.
  \label{eq:evoldis}
\end{equation}
We center the derivatives at the point $(n+1/2, i-1/2)$, as dictated by
the DA scheme and obtain ${g_{,u}|}^{n+1/2}_{i-1/2}$ by means of an iterative procedure.
In the first iteration we set $g_{,u}=g_{,u}|^{n+1/2}_{i-1}$ and get a first
approximation of $g^{n+1}_{i}$ via the evolution equation. Then, we use
this value to obtain a guess for ${g_{,u}|}^{n+1/2}_{i-1/2}$ which is
then used to get a better approximation for $g^{n+1}_{i}$. This
procedure is repeated a sufficient number of times to ensure
convergence. 

Unfortunately, when solving a $3$ dimensional problem, the computational
requirements of integrating from the origin ($r=0$) are formidable. However, it
is possible to start the integration from a finite
value of $r$, assuming consistent values of $\beta$, $U$, $Q$, $V$ and
$J$ are known on this boundary (which is refered to as the worldtube
boundary) as well as the value of $J$ on an initial hypersurface~\cite{wintam}. 

To illustrate the usefulness of the presented algorithm, we apply it to
model ({\it i}) the propagation of linear waves on a Minkowski background and
({\it ii}) the problem of scattering off a Schwarzschild black hole in 3 dimensions.

\subsection{Linear Waves on a Minkowski background}

In the past, analytical solutions of linearized Einstein equations (in
the characteristic formulation) have been found which describe waves
propagating on a flat background~\cite{cce}. These solutions
provide an important test bed for the algorithm, since the numerically
obtained solutions must converge to the analytic values given by
\begin{equation}
\beta =0, \, 
V = r 
\end{equation}
with $J$ and $U$ obtained from a solution ($\Psi$) of the
scalar wave equation by 
\begin{eqnarray}
J_{,r} &=&  \frac{(r^2 \eth^2 \Psi)_{,r}}{2 r^2} \, ,\\
U_{,r} &=& -2 \frac{\eth( \eth \bar \eth \Psi + 2 \Psi)}{r^2} \, .
\end{eqnarray}  

In order to test the algorithm, we choose a solution of the wave
equation in $3$ dimensions, that represents an outgoing wave with angular
momentum $0 \le l \le 6$ of the form
\begin{equation}
 \Phi =(\partial_z)^6\frac{\alpha}{u^2r} \, , \label{eq:exacsol}
\end{equation}
where $\partial_z$ is the $z$-translation operator. The resulting
solution is well behaved above the singular light cone $u=0$.

Choosing initial data of very small amplitude $( \alpha \approx
10^{-9})$; we used these solutions to give data at $u=1$ (with the
inner boundary set at $R=1.5$) and compared the numerical and exact
solutions over time for different values of $\epsilon$. The computation
was performed on grids of size $N_x$ equal to $41$, $53$, $65$ (with
the number of angular points $N_\xi= (N_x-1)/2 + 5, $ and  the ratio
$\Delta u/(4 \Delta r) = 1/8$). The $L_2$ norm of the error was
calculated over the entire grid and plotted against time for different
values of the dissipation parameter. Figure \ref{errwav} shows the
logarithm of the error in $J$ vs. time (for runs with $N_x=65$).
For $\epsilon=0$ the evolution is unstable, as can be seen by
the exponential growth of the error. For $\epsilon=0.005$ the 
instability appears at a later time, also with an exponential
growth. However, for $\epsilon=0.05$ the run proceeds stably and the
error remains under control. It is important to note that the magnitude
of the dissipation needed to achieve a stable run is very small and
therefore the ``damping'' of the solution in not severe.

\subsection{Nonlinear scattering off a Schwarzschild black hole}
\label{nonlinewav}
The characteristic initial value problem on an outgoing null
hypersurface requires inner boundary conditions on the worldtube.
Here we consider an example in which the inner boundary $\Gamma$
consists of an ingoing nullcone (see figure~\ref{fig:pulse}). We adopt
coordinates $x^A$ which follow the ingoing null geodesics and foliate
$\Gamma$ (chosen to correspond to ingoing $r=2m$ surface
in a Schwarzschild spacetime) by slices separated by constant parameter $u$. 
In these coordinates, the Schwarzschild line element takes form
\begin{eqnarray}
   ds^2=-\left( 1 - {2m \over r} \right)du^2
        -2 dudr
        +r^2q_{AB}dx^Adx^B.
\end{eqnarray}
The initial data correspond to setting $J=0$ as data on $u=0$, with
the boundary conditions $\beta=U=Q=0$ and $V = r -2 m$ on
$\Gamma$.

We pose the nonlinear problem of gravitational wave scattering off a
Schwarzschild black hole by retaining these boundary conditions on
$\Gamma$ but we choose null data at $u=0$ corresponding to an incoming
pulse with compact support\footnote{Recall that the data on
the initial hypersurface can be chosen freely in the characteristic
formulation~\cite{jeffnew}.},
\begin{equation}
   J(u=0,r,x^A) = \left\{ \begin{array}{ll}
  \displaystyle{\lambda \left(1 - \frac{R_a}{r} \right)^4 \,
                        \left(1 - \frac{R_b}{r} \right)^4 \;
                        \sqrt{\frac{4 \pi}{2 l + 1}} \; {}_{2} Y_{l,m}}
                    & \mbox{if $r  \in [R_a,R_b]$} \\
 		    & \\
                  0 & \mbox{otherwise,}
                        \end{array}
                        \right.
\end{equation}
where ${}_{2}Y_{l,m}$ is the spin-two spherical harmonic~\cite{goldberg}. 

The code was run for different values of $\lambda$ under different
choices of the dissipation parameter. In all cases, unstable evolutions
resulted from the choice $\epsilon=0$, however for nonzero values of $\epsilon$ the
code ran without any stability problem as illustrated in
figure~\ref{fig:evol1} (for a run where $\lambda=1, l=2, m=0$). 

Yet, as expected of any dissipative algorithm, the solution decreases
in amplitude with time.  This highlights the need to carefully tune the
value of $\epsilon$.  Notwithstanding this fact, it is important to
stress once again that this set of runs would not have been possible
without dissipation.

This problem was originally studied in the perturbative regime by
Price~\cite{price}. There is no known analytic solution to the problem
in the nonlinear regime and applying numerical methods is the only way
to study it. The accuracy of the dissipative scheme can be assessed
indirectly by inspection of the gravitational waves produced by the
system. Gravitational waves can be described in terms of two
{\it polarization modes} (refered to as {\it  plus} and {\it
cross} modes)~\cite{mtw}. However, when considering spacetimes with axisymmetry,
the cross mode must vanish and this fact can be used to test the
algorithm. Calculating the gravitational radiation is a rather involved
problem that exceeds the scope of this work. A set of algorithms to
numerically calculate the gravitational wave forms was constructed in
the characteristic formulation in~\cite{hpgn} and tested under
different situations. We used these algorithms in the present work to
calculate the polarization modes for the choice $ \epsilon = \Delta r$
and with an axisymmetric pulse with $l=2, m=0$ as the initial data. The
cross polarization mode actually converges to zero in second order indicating an
accurate discretization of Einstein equations, as can be seen in Figure
\ref{fig:conver}.

\section{Conclusion}
The algorithm described in this work represents a valuable tool for the
study of nonlinear problems in the characteristic formulation. Its use
enables long term evolution that would otherwise be impossible. Yet, there
is still much room for improvement as the number of numerical techniques 
adapted to characteristic type evolutions is scarce (as opposed to
the situtation in the Cauchy type evolution where one has at hand
a great number of algorithms). The variety of physical problems where
propagating waves are to be described, highlights the need of further 
investigations on ``characteristic'' algorithms.

Of particular interest is the application of these type of
algorithms to the characteristic module
constructed to model the collision of a binary black hole
self-gravitating system. In this problem, it is imperative to have
robust enough schemes capable of dealing with highly nonlinear
fields. The complexity of the problem inspired the creation of the
Binary Black Hole Grand Challenge Alliance, where a group of U.S.
universities and outside collaborators are joining efforts to tackle
the problem~\cite{all}.  A strategy to study this problem is a
``hybrid'' scheme that implements at the same time a Cauchy evolution
(for the region near the black holes) and a Characteristic evolution
(for the exterior region).  This approach is called {\it
Cauchy-characteristic matching (CCM)}~\cite{Bis,manual,dinv}, and in
principle, its implementation manages to avoid the problems and to
exploit the best features of each evolution scheme. CCM has been shown
to work (and outperform traditional outer boundary conditions) 
in situations where special symmetries were
assumed~\cite{jcp,excision} and its full $3$ dimensional application in G.R. is
currently under study.  The characteristic code is one of the pieces of
this bigger algorithm and the need for robust performance prompted
this investigation.  However, its use is not limited to G.R. Any
hyperbolic system describing waves will have an evolution equation
similar to Eq. (\ref{weqnul}). The
algorithm presented in this work should provide a useful tool in the
numerical modeling of these systems.

\section{Acknowledgments}
This work has been supported in part by the Andrew Mellon
Fellowship, by the NSF grants PHY 9510895 and NSF INT 9515257 to the
University of Pittsburgh and by the Binary Black Hole Grand Challenge
Alliance, NSF grant PHY/ASC 9318152 (ARPA supplemented).
 The author would like to thank Jeffrey Winicour, Roberto G\'omez and
Nigel Bishop for
valuable suggestions, also to the Universities of South Africa and of
Durban-Westville for their hospitality in completing part of this work.
Computer time for this project has been provided by the Pittsburgh
Supercomputing Center under grant PHY860023P.

\begin{figure}
\caption{Domains of dependence in the Cauchy and characteristic 
initial value problems.}
\label{fig:domdep}
\end{figure}

\begin{figure}
\caption{The characteristic scheme for the exterior problem. Initial data
is given on ${\cal N}_o$ and boundary data on $\Gamma$.}
\label{fig:worldtube}
\end{figure}

\begin{figure}
\caption{Plots of $|S|$ corresponding to different choices of $\epsilon$ and $\lambda$.
A $(\lambda=0, \epsilon=0$); B $(\lambda=0.02, \epsilon=0$);
C, $(\lambda=0.02, \epsilon=0.02$) and D, $(\lambda=0.02, \epsilon=0.2$) which illustrates
how adding artificial dissipation ensures stability. However, as can be seen
in D for a high value of $\epsilon$ the damping of the high frequency modes
might be severe.}
\label{vareps}
\end{figure}

\begin{figure}
\caption{The logarithm of $|E| \equiv |J^{num}-J^{anal}|$ (the numerical
and analytic values of $J$) is shown for different values of $\epsilon$. For
$\epsilon=0.05$ the evolution is stable as opposed to the unstable
evolutions correspondent to $\epsilon=0$ and $0.005$.}
\label{errwav}
\end{figure}

\begin{figure}
\caption{Scattering off a Schwarzschild black hole. The bold dashed line
illustrates the incoming pulse.}
\label{fig:pulse}
\end{figure}

\begin{figure}
\caption{Plots of the field variable $J$ at a representative angle vs.
a compactified radial coordinate $x=r/(1+r)$. The value of the mass is
$m=0.5$, the amplitude of the initial pulse is $\lambda = 5$, $R_a =
4$ and $R_b=8$. The runs correspond to different choices of $\epsilon$.
The solid lines indicates the initial data at $u=1$. Box (a) shows the
run for $\epsilon=0$, after a short time the obtained values diverge;
(b) corresponds to the choice $\epsilon=0.005$ showing a run that
neither show signs of instability nor much damping of the pulse; (c) in
turn corresponds to $\epsilon=0.02$, although there is no sign of
instability the solution has been damped considerably.}
\label{fig:evol1}
\end{figure}

\begin{figure}
\caption{Convergence of the cross polarization mode to zero (in these
runs $\epsilon$ was chosen equal to $\Delta r$). The slope is $1.99$,
confirming second-order accuracy of the obtained wave form.}
\label{fig:conver}
\end{figure}

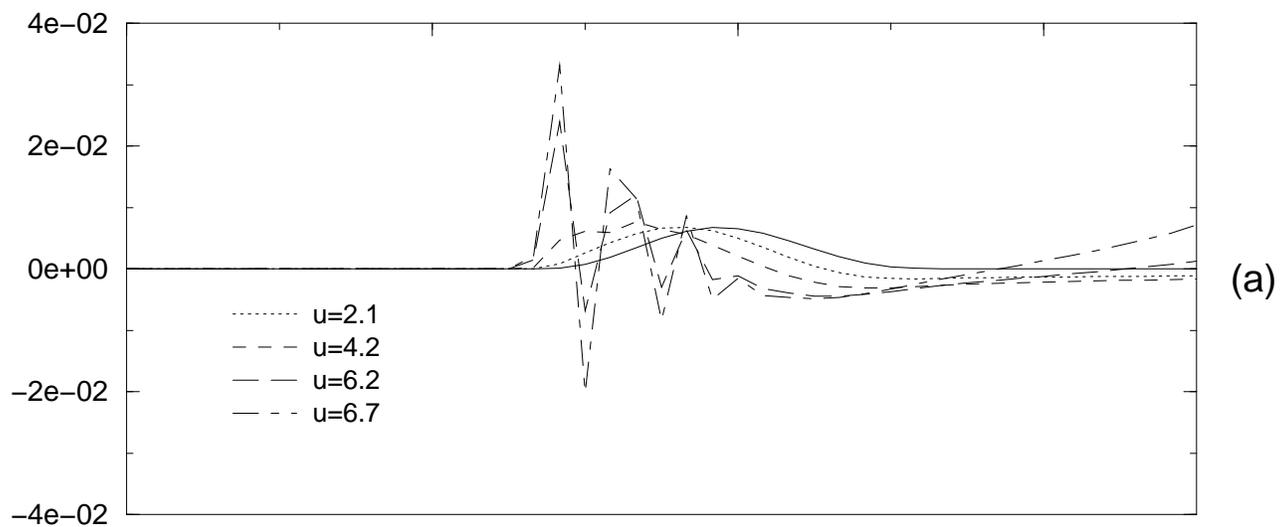
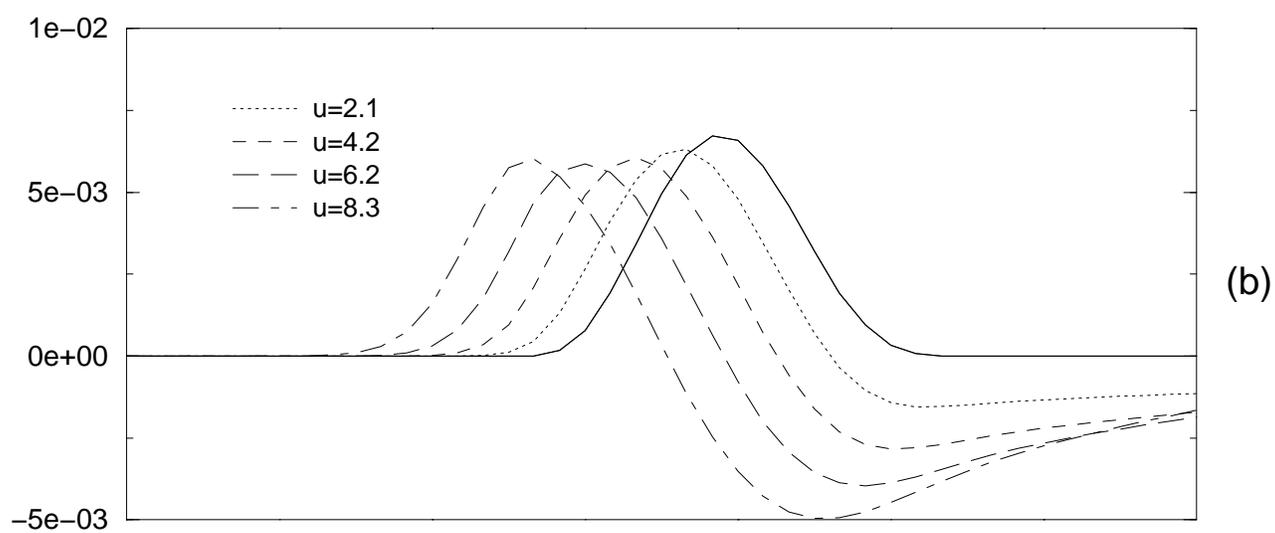
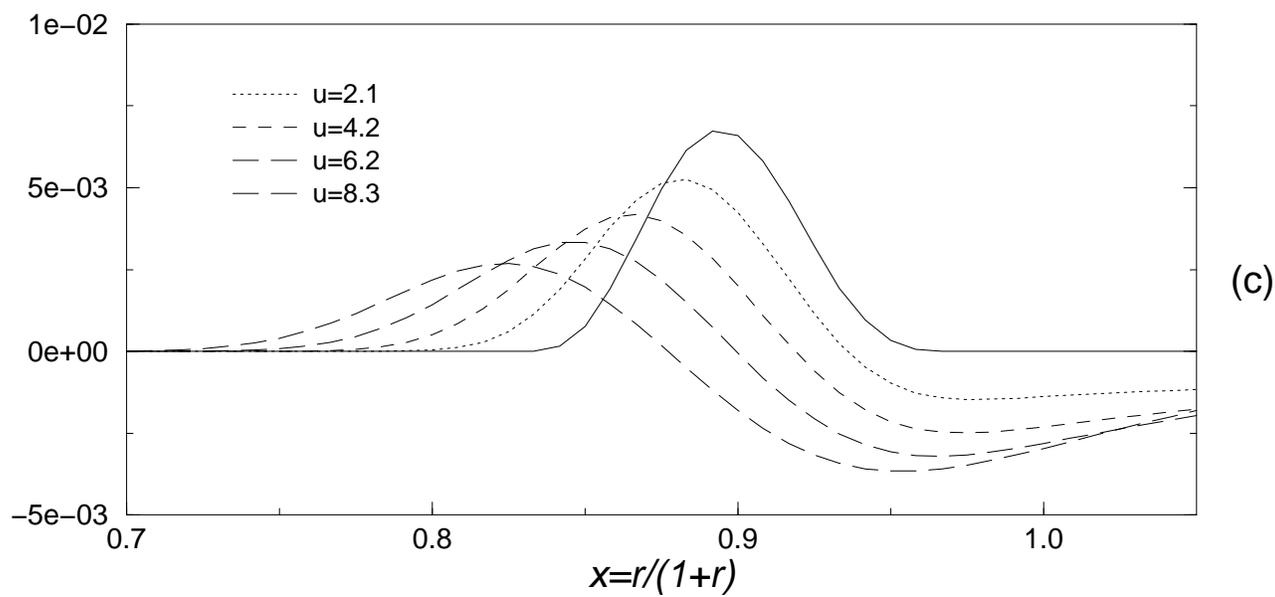
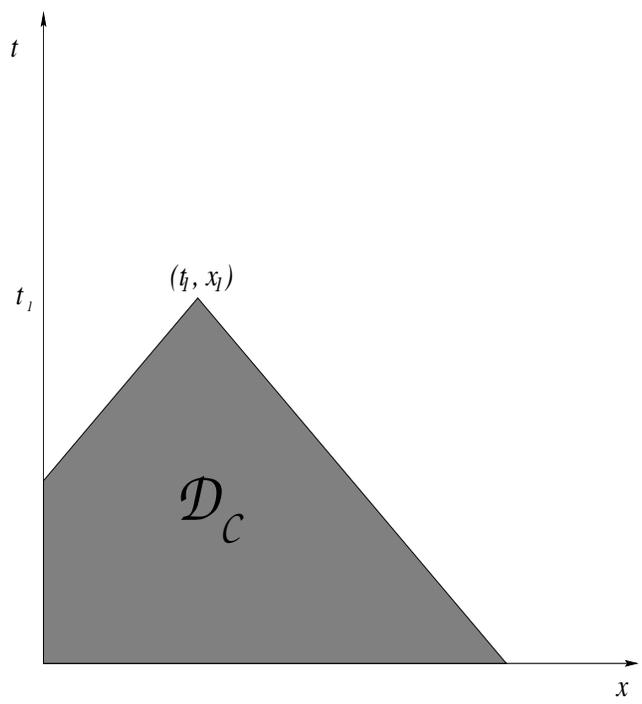
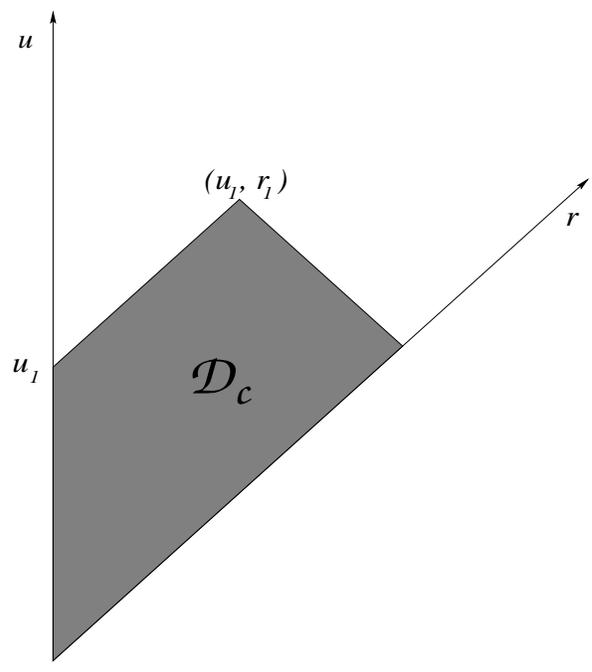
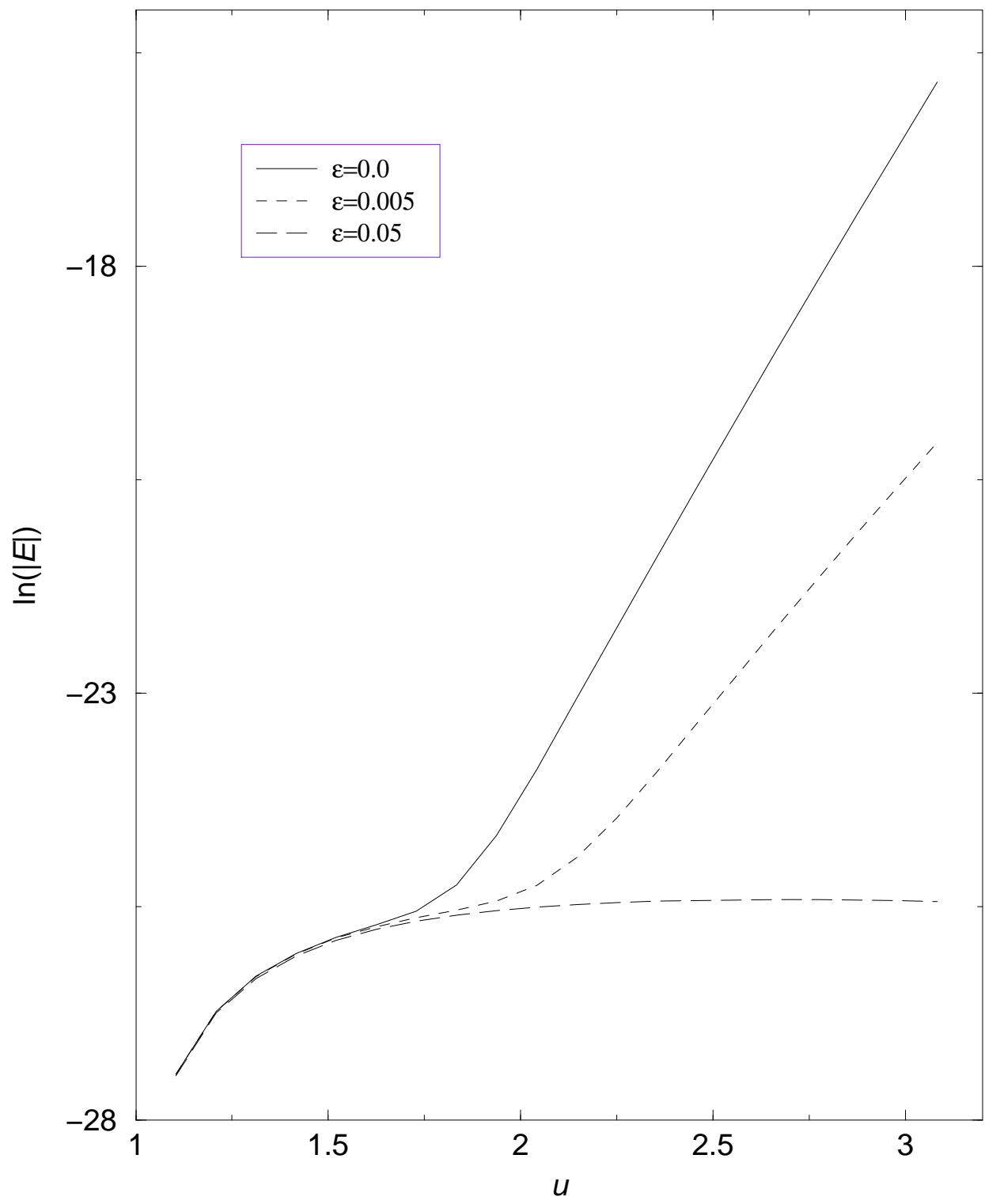
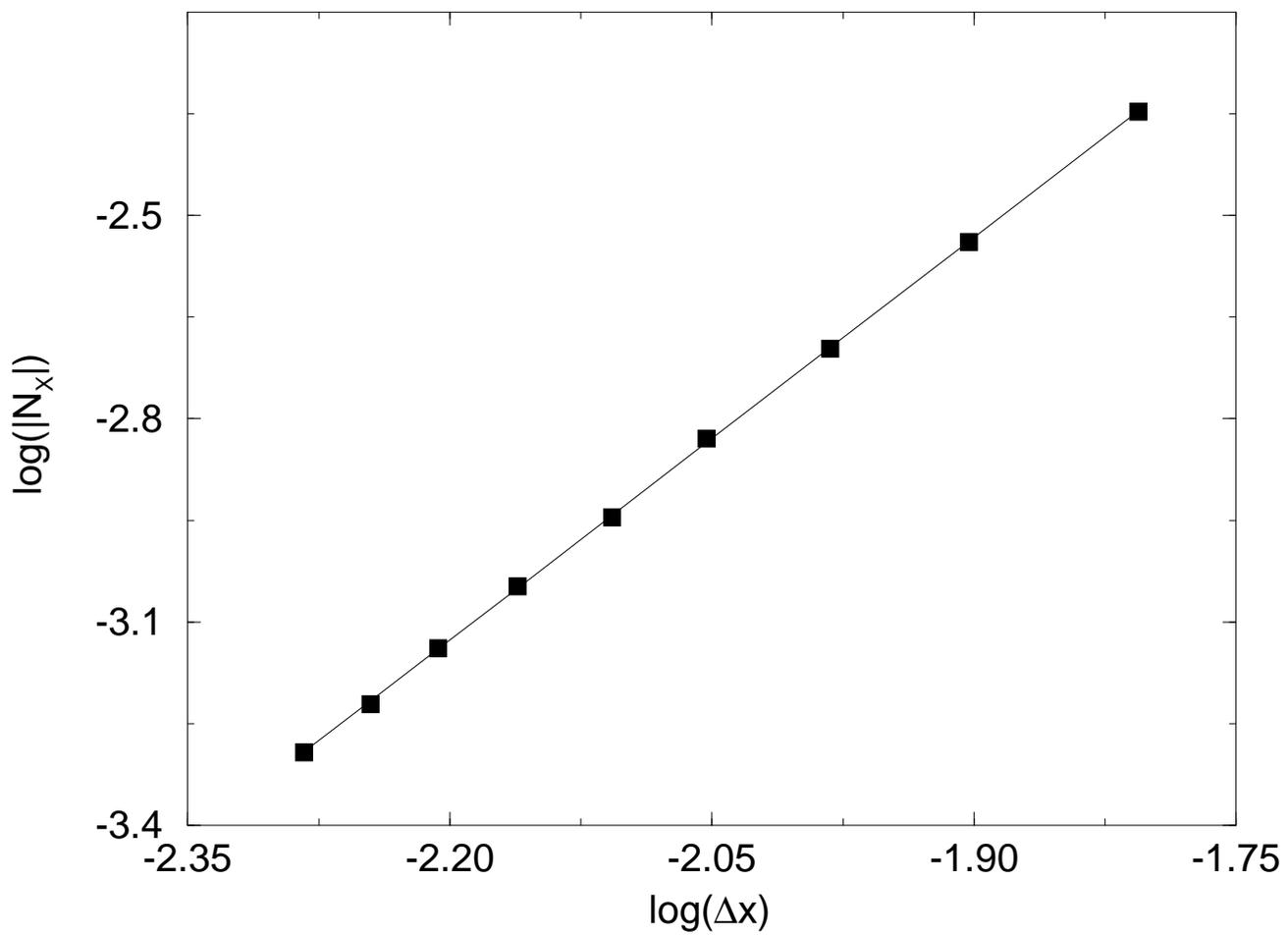
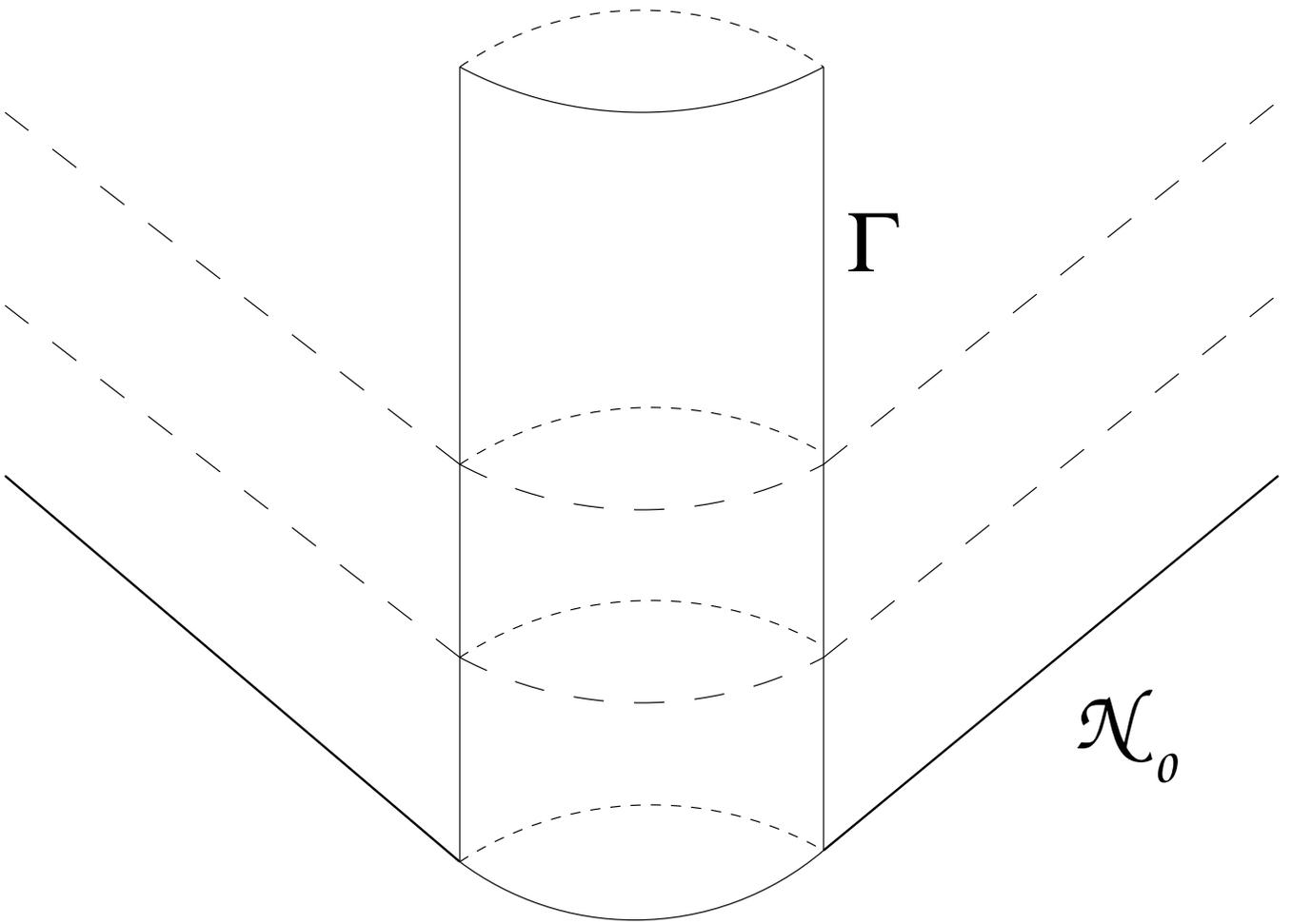
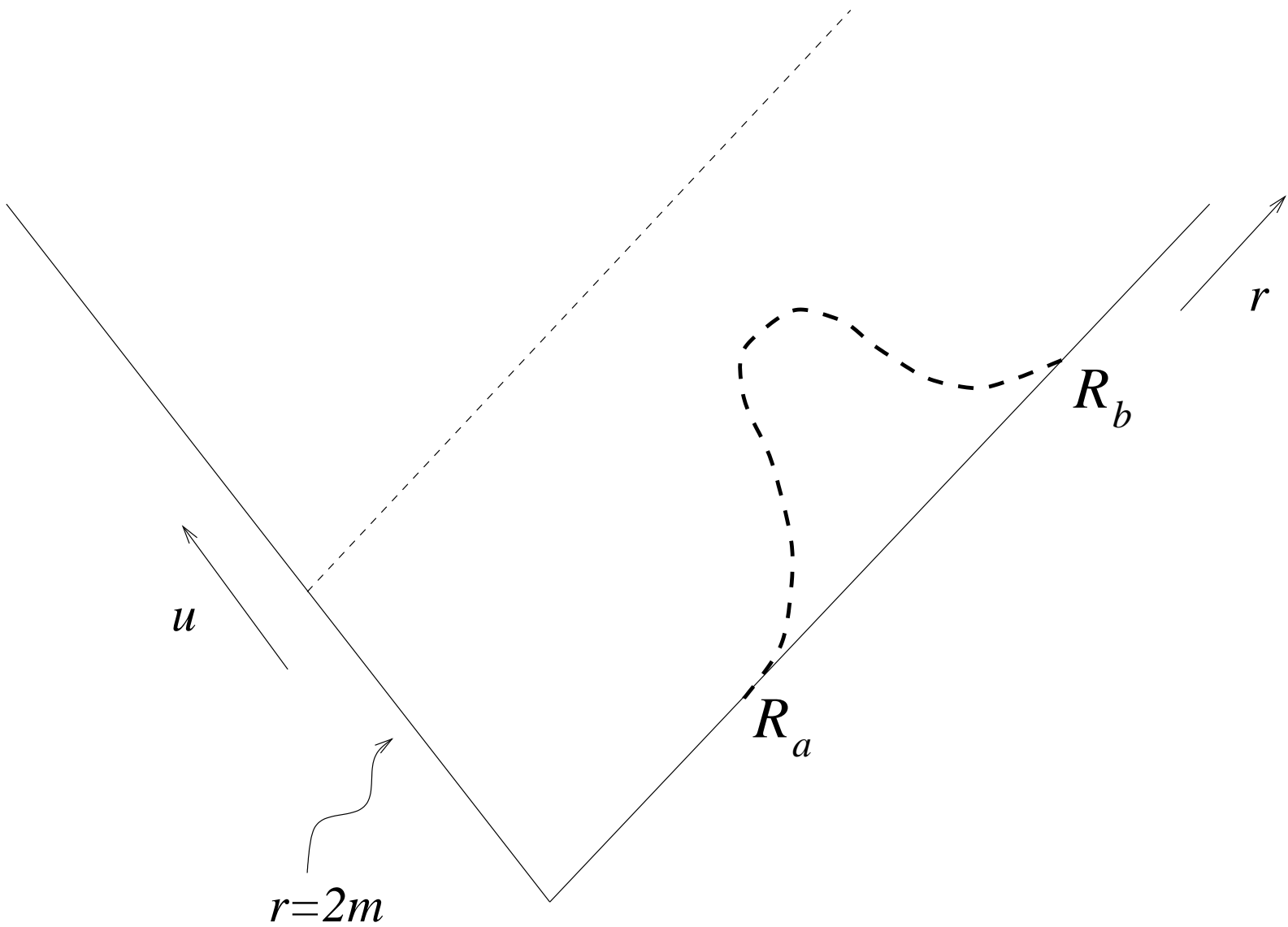
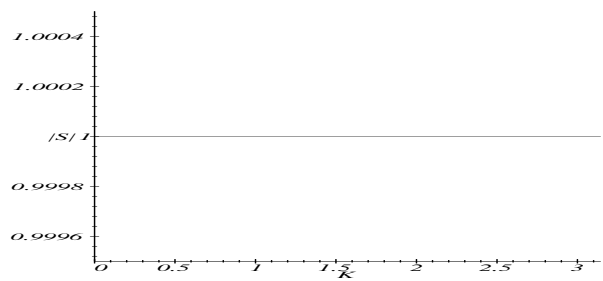
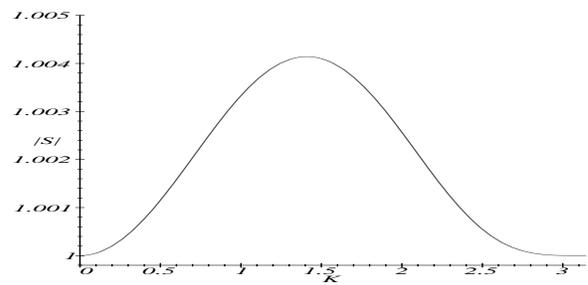
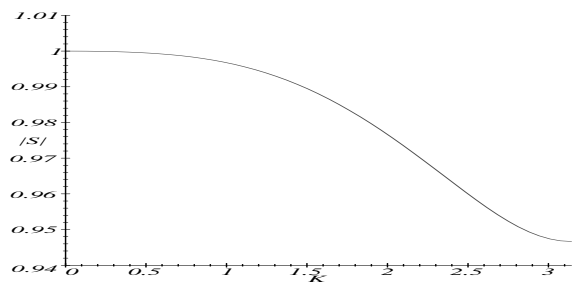
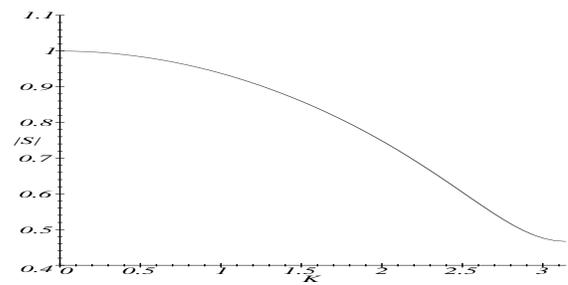
\end{document}